\documentstyle[prl,multicol,psfig,aps]{revtex}
\begin{document}
\draft


\title{Infrared absorption from Charge Density Waves in magnetic manganites}
\author{P. Calvani, G. De Marzi, P. Dore, S. Lupi, P. Maselli, F. 
D'Amore, and S. Gagliardi} 
\address{Istituto Nazionale di Fisica della Materia} 
\address{and Dipartimento di Fisica,
Universit\`a di Roma ``La Sapienza'', Piazzale A. Moro 2, I-00185 Roma, Italy}
\author{S-W. Cheong} 
\address{AT\&T Bell Laboratories, Murray Hill, New Jersey 07974, U.S.A.}    
\address{and Department of Physics, Rutgers University, Piscataway, New Jersey
08855, U. S. A.}

\date{\today}

\maketitle
\begin{abstract} 
The infrared absorption of charge density waves coupled to a magnetic 
background is first observed in two manganites La$_{1-x}$Ca$_{x}$MnO$_3$ with 
$x = 0.5$ and $x = 0.67$. In both cases a BCS-like gap $2\Delta (T)$,
which for $x=0.5$ follows the hysteretic ferro-antiferromagnetic transition,
fully opens at a finite  $T_0 < T_{Neel}$, with $2\Delta(T_0)/k_BT_c \simeq 5$. 
These results may also explain the unusual coexistence of charge ordering and 
ferromagnetism in La$_{0.5}$Ca$_{0.5}$MnO$_3$.
\end{abstract}

\begin{multicols}{2}
\narrowtext

The close interplay between transport properties and magnetic ordering in the 
colossal magnetoresistance (CMR) manganites 
La(Nd)$_{1-x}$Ca(Sr)$_{x}$MnO$_3$ is presently explained in terms of 
magnetic double exchange promoted by polaronic carriers along the path
Mn$^{+3}$ - O$^{-2}$ - Mn$^{+4}$.\cite{Millis}  
Charge hopping promotes the alignment of Mn$^{+3}$ and Mn$^{+4}$ 
magnetic moments, and vice versa. The polaronic effects are due to the dynamic 
Jahn-Teller distortion of the oxygen 
octahedra around the Mn$^{+3}$ ions. The above mechanism explains how, in 
manganites with $0.2 < x < 0.48$, any increase in the magnetization enhances 
the dc conductivity, and vice versa. 
However, La$_{0.5}$Ca$_{0.5}$MnO$_3$ shows an unpredicted 
coexistence of ferromagnetism and incommensurate charge ordering (CO). 
This compound is paramagnetic at room temperature,
becomes ferromagnetic (FM) at $T_c \simeq$ 225 K and, by further cooling (C),
antiferromagnetic (AFM) at a N\'eel temperature  
$T_N^C \simeq$ 155 K.\cite{Radaelli} Upon heating the sample (H) the FM-AFM 
transition is instead observed at $T_N^H \simeq$ 190 K .\cite{Schiffer} 
The dc conductivity $\sigma (0)$ of La$_{0.5}$Ca$_{0.5}$MnO$_3$ is quite 
insensitive to the PM-FM transition at $T_c$.\cite{Schiffer}  
X-ray, neutron\cite{Radaelli2} and electron diffraction\cite{Chen} show 
quasi-commensurate charge and orbital ordering in the AFM phase with wavevector 
$\vec q = (2 \pi/a)({1\over 2}-\epsilon, 0, 0)$. The incommensurability 
$\epsilon$ increases with temperature and
follows the hysteretic behavior of the AFM-FM transition, until charge ordering
disappears above the Curie point $T_c$.\cite{Chen} At higher Ca doping, for 
$0.5 \alt x \alt 0.75$, a transition to a charge ordered 
phase\cite{Ramirez} is observed in the paramagnetic phase at $T_{CO}$. 
$T_{CO}$ is a maximum (265 K) for $x = 0.67 \simeq {2 \over 3}$, 
where the charge ordering is commensurate with the lattice. Below $T_{CO}$, the
system enters at $T_N$ an antiferromagnetic phase. For $x = 0.67$, 
$T_N \simeq 140$ K.

In the present paper, the unexpected coexistence of incommensurate charge 
ordering and ferromagnetism in La$_{0.5}$Ca$_{0.5}$MnO$_3$ is investigated 
by infrared spectroscopy. The spectra of stoichiometric 
LaMnO$_3$ and of La$_{0.33}$Ca$_{0.67}$MnO$_3$ are also examined for 
comparison. 
Both doped manganites allow to study the optical response of charge
density waves (CDW) interacting with a magnetic background. Such an experiment
could hardly be done in the low-dimensional systems where CDW 
are usually observed.


The present infrared spectra have been collected on the same 
La$_{1-x}$Ca$_{x}$MnO$_3$ samples, prepared as 
described in Ref. \onlinecite{Schiffer}, where the above diffraction 
studies\cite{Chen,Ramirez} have been performed.  
The oxygen stoichiometry has been accurately controlled,\cite{Chen} to fulfill 
doping requirements which are particularly strict at $x = 0.5$. The sintherized 
compound has been 
finely milled, diluted in CsI (1:100 in weight), and pressed into pellets 
under vacuum. One thus obtains reliable data 
when both the reflectance and the transmittance of the sample are 
too low, as shown in similar experiments\cite{rc96} on 
La$_{2-x}$Sr$_{x}$NiO$_4$ and Sr$_{2-x}$La$_{x}$MnO$_4$. The infrared intensity 
$I_{s}$  transmitted by the pellet containing the oxide and that, $I_{CsI}$, 
transmitted by a pure CsI pellet, have been measured at the same 
$T$. One thus obtains
a normalized optical density that, as shown in Ref. \onlinecite{physb98},
is proportional to the optical conductivity $\sigma (\omega)$ of 
the pure perovskite, over the frequency range of interest here:

\begin{equation}
O_d(\omega) = ln[I_{CsI}(\omega)/I_s(\omega)] \propto \sigma({\omega})
\end{equation}

The spectra have been collected by two interferometers 
between 130 and 10000 cm$^{-1}$, and by accurately thermoregulating 
the samples within $\pm$ 2 K between 300 and 20 K. The susceptibility 
$\chi (T)$ of La$_{0.5}$Ca$_{0.5}$MnO$_3$ has been measured at zero static 
field in a commercial apparatus by the Hartshorn method, at a frequency of 
127 Hz.   


The optical density $O_d(\omega)$ of La$_{0.5}$Ca$_{0.5}$MnO$_3$ 
is compared in Fig. 1 with that of stoichiometric LaMnO$_3$.
The latter compound  shows in Fig. 1(a) a negligible absorption in the 
midinfrared at all temperatures. At least 10 phonon lines, 
out of the 19 predicted\cite{Couzi} under the orthorombic symmetry 
of LaMnO$_3$, are also found out in Fig. 1(a) by a fit to Lorentians. 
A detailed analysis of the 
phonon spectrum of this family of manganites will be reported elsewhere. 

\begin{figure}
{\hbox{\psfig{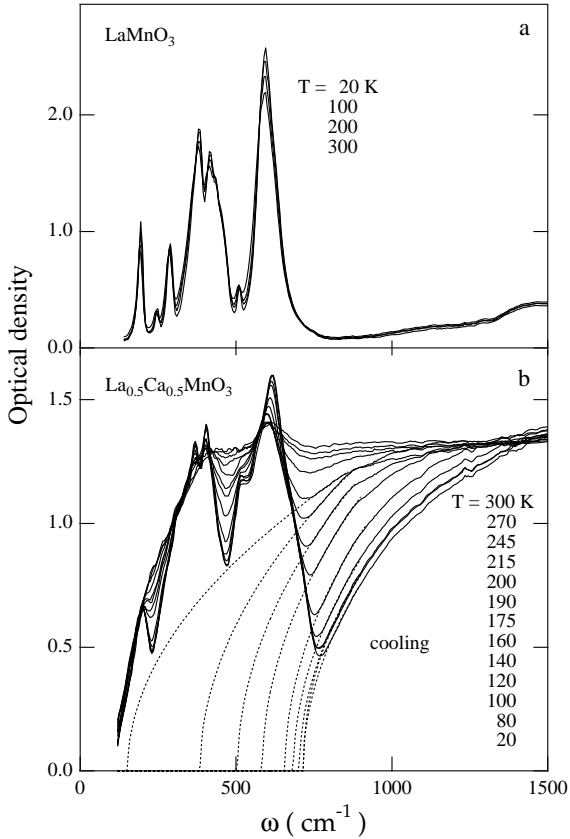}}}
\caption{The optical density $O_d$  of polycrystalline LaMnO$_3$ (a) 
and La$_{0.5}$Ca$_{0.5}$MnO$_3$ (b) as the sample is 
cooled from 300 K to 20 K. The dashed lines in (b) represent extrapolations 
based on Eq. (3).} 
\label{fig1}
\end{figure} 

Let us now consider Fig. 1(b), where the optical density of 
La$_{0.5}$Ca$_{0.5}$MnO$_3$ is reported for decreasing temperatures. At 300 K 
the phonon peaks are shielded by a broad background which, after a
straightforward comparison with Fig. 1(a), can be attributed
to the mobile\cite{nota} holes introduced by Ca doping. However, at a $T$  
between 245 and 215 K, a minimum appears in that background, which 
gradually deepens by further lowering the temperature. This signature of an
optical gap provides evidence for increasing localisation of the carriers. 
Indeed, as $T$ lowers, phonon peaks similar to those in Fig. 1(a) become 
more and more evident, showing that the screening action of the mobile holes  
weakens. Moreover, the states lost in the gap region for decreasing 
temperature are transfered to higher frequencies
(see the crossing point of the absorption curves at $\simeq$ 1500 
cm$^{-1}$), not to a coherent peak at $\omega = 0$. This finding is consistent
with the regular decrease in $\sigma (0)$\cite{Schiffer} 
observed in La$_{0.5}$Ca$_{0.5}$MnO$_3$ between room temperature and $T_N$. 

A loss of spectral weight such as in Fig. 1(b) can be measured by the
variation of an effective number of carriers $n_{eff} (T)$, obtained by 
integrating $\sigma (\omega)$ between two suitable frequencies. 
By taking into account Eq. (1), we can define for the present purpose a
quantity: 

\begin{equation}
n^*_{eff}(T) = \int_{\omega_1}^{\omega_2} O_d(\omega)\,d\omega
\propto n_{eff}(T) 
\end{equation}

\noindent
As suitable integration limits one may take the two frequencies where 
$O_d (\omega)$ does not appreciably change with temperature, $\omega_1$ = 
200 cm$^{-1}$ and $\omega_2$ = 1500 cm$^{-1}$.  
The values thus obtained for $n^*_{eff} (T)$ along a thermal cycle are plotted 
in Fig. 2(a). The inset compares the spectra observed at the same $T$ = 160 K 
upon cooling and heating the sample. When the latter is cooled 
from $T \agt T_c$, spectral weight is increasingly lost in the far infrared. 
A comparison with the corresponding curves of the magnetic susceptibility 
$\chi (T)$ in Fig. 2(b) shows that the decrease in $n^*_{eff} (T)$ is completed 
well below the N\'eel temperature $T_N^C$. Full localisation of the carriers 
is then observed only when $\chi (T)$ has reached its minimum value and an AFM 
phase is established in the whole sample.  When heating the sample, 
$n^*_{eff} (T)$ starts increasing around 120 K, again well below $T_N^H$. 

The behavior of the infrared absorption in Fig. 1(b) and 2(a) in a temperature
range where electron diffraction observes incommensurate charge ordering,
points toward the formation a charge density wave. One can then extract 
from Fig. 1(b) the optical gap $2\Delta (T)$ and
compare its behavior with that expected for a CDW. In that case, 
$\sigma(\omega)$ at $\omega \agt 2\Delta (T)$ is similar to that of a
semiconductor in the presence of direct band to band transitions.\cite{Lee} By 
remembering again Eq. (1), one can then fit to the experimental
$O_d (\omega)$, for $\omega \agt 2\Delta (T)$, the expression: 
                           
\begin{equation}
\sigma(\omega) \propto [\omega - 2\Delta(T))]^{\alpha} .
\end{equation}

\noindent
The curves thus obtained, reported as dashed lines in Fig. 1(b), well 
describe the resolved part of the gap profile at all temperatures with 
the same $\alpha = {1\over 2}$. The resulting values for 
$\Delta (T)$ are plotted in Fig. 2(c) as full circles, those corresponding to 
heating the sample as open circles. BCS-like fits for $2\Delta(T)$ have been 
successfully applied to the optical behavior of charge density waves in polar 
systems,\cite{Katsufuji} even if the resulting values for $2\Delta_0/k_BT_c$ 
are much higher than that (3.52) expected for a weak electron-lattice coupling.
An analytic expression for $2\Delta (T)$ between $T$ = 0 and $T_c$,\cite{Burns} 
which also holds under moderately strong coupling,\cite{Thouless} can be 
written as: 

\begin{equation}
\Delta (T)/\Delta_0 = tanh \Biggl [{\Delta (T) T_c \over 
\Delta_0 T} \Biggr ]
\end{equation}

\begin{figure}
{\hbox{\psfig{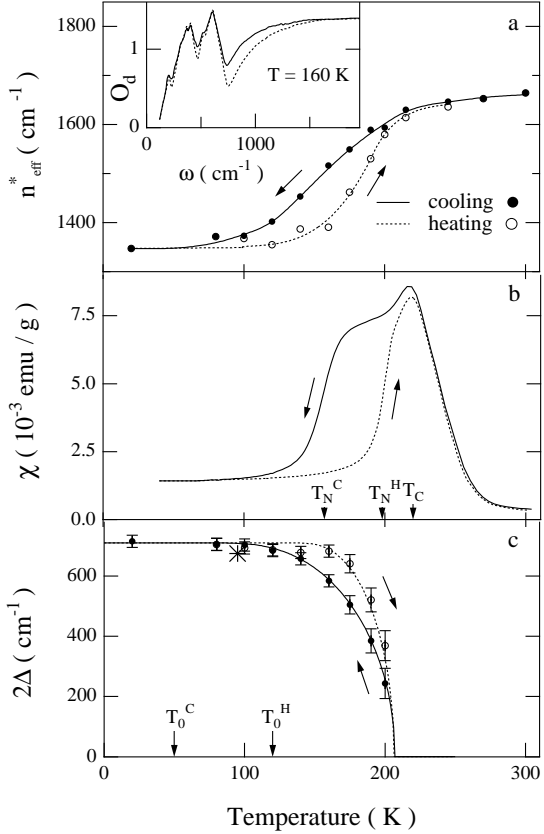}}}
\caption{Effective number of carriers $n^*_{eff} (T)$ (a), zero-field magnetic 
susceptibility $\chi (T)$ (b), and width of the optical gap $2\Delta$ (c) 
vs. temperature in La$_{0.5}$Ca$_{0.5}$MnO$_3$, as the sample is cooled (full
circles, solid lines) or heated (open circles, dashed lines). The lines in (a)
are guides to the eye, those in (c) are fits to Eq. (5). The inset in (a) shows
the optical densities measured at the same $T = 160$ K by cooling (solid) 
and heating (dashed). In (b), the $T_N^C$ and $T_N^H$ values here found at a
static field $H = 0$ are slightly higher than reported in Ref. [3] for $H$ = 
1 T. In (c), the star corresponds to the activation energy extracted from the 
dc resistivity of Ref. [3] (see text).}
\label{fig2}
\end{figure} 

\noindent
Unlike for an ordinary CDW, here the antiferromagnetic background fully 
localizes the charges at a finite $T_0 <T_N$, which moreover depends on 
the thermal procedure. We find that this effects can be 
simply taken into account by replacing in Eq. (4) $T = 0$ by $T_0^X (X=C,H)$: 

\begin{equation}
\Delta (T)/\Delta_0 = tanh \Biggl [{\Delta (T) (T_c-T_0^X) \over 
\Delta_0 (T-T_0^X)} \Biggr ]
\end{equation}
 
\noindent
Indeed in Fig. 2(c), Eq. (5) well fits both series of data by using the same 
values $2\Delta_0 = 710 \pm 20$ cm$^{-1}$ and $T_c \simeq$ 210 K, which gives 
$2\Delta_0/k_BT_c \simeq$ 4.8. $T_0^C$ = 50 K and $T_0^H$ = 120 K are  
introduced into Eq. (5) for cooling and heating, respectively.
Any attempt to raise either $T_0$ value towards the corresponding $T_N$ 
cosiderably worsen the fit, thus confirming that the gap fully opens well below 
the N\'eel temperature. One may notice that the dc resistivity reported 
for La$_{0.5}$Ca$_{0.5}$MnO$_3$ in  Ref. \onlinecite{Schiffer} follows an
exponential behavior $\rho (T) \propto$ exp$[E_0/k_BT]$ below $\sim$ 110 K. The
activation energy $E_0 = 675 \pm$ 25 cm$^{-1}$ (star in Fig. 2(c))
is in excellent agreement with the present value of 
$2\Delta_0$. A refinement of the present 
infrared data would instead be needed in order to investigate the nature of 
the gap (or possibly pseudogap) slightly above $T_c$. 

The results of Fig. 2 can be usefully discussed in connection with those of
electron diffraction in the same La$_{0.5}$Ca$_{0.5}$MnO$_3$ powder.
The behavior of the incommensurability parameter $\epsilon$, reported in Fig. 
2 of Ref. \onlinecite{Chen}, is 
quite similar to that of $n^*_{eff} (T)$ in Fig. 2(a). At 240 K, where 
$n^*_{eff} (T)$ starts decreasing, weak peaks from a charged superlattice 
appear in the electron diffraction spectra, corresponding to 
$\epsilon \simeq$ 0.10. As $T$ lowers, $\epsilon (T)$ decreases to
0.01 at $T = T_N$ and vanishes below $\sim$ 120 K. Along a thermal cycle, 
$\epsilon (T)$ follows a hysteretic behavior quite similar to that 
of $n^*_{eff} (T)$ in Fig. 2 (a). Such results, and the present 
observation of $2\Delta (T)$, can be attributed either to an 
incommensurate, homogeneous charge density wave, or to discommensurations 
between ordered domains which are commensurate with the lattice. The former 
assumption can be ruled out at $x = 0.5$ and in the presence of a strong
electron-lattice coupling which creates polaronic carriers. The latter 
possibility points toward a phase separation scenario, as discussed below.

The average dimension of the commensurate domains would 
be\cite{Chen} $L = a/p \epsilon$, where $a \simeq 0.55$ nm is the lattice 
parameter in the Mn-O planes, and $p =2$ is the order of 
commensuration. This gives $L \simeq 5a$ at $T_c$, $L \sim 50a$ at $T_N$. 
One may expect that an AFM phase is established in such regions, as observed 
on a macroscopic scale below $T_{CO}$ for $x > 0.5$.\cite{Ramirez}
If this is true, the ferromagnetism observed in La$_{0.5}$Ca$_{0.5}$MnO$_3$
below $T_c$ can only take place in the disordered regions which separate the 
commensurate AFM clusters. Even there however, the charges should 
have a reduced mobility. 
To understand what may happen, one should consider that single-particle 
hopping undergoes intrinsical limitations at $x \ge 0.5$. Indeed, 
the dynamical distribution of Mn$^{+3}$ and Mn$^{+4}$ ions in the lattice
is not purely statistical, as the system reduces its free energy 
by alternating distorted and undistorted octahedra. Then for $x = 0.5$,
a valence electron on Mn$^{+3}$ is surrounded by Mn$^{+4}$ 
nearest neighbors and may easily establish with these latters a double 
exchange mechanism. However, the same electron after the first jump will be 
surrounded by Mn$^{+3}$ ions (except for the starting site, now Mn$^{+4}$) so 
that a second jump in some forward direction is prevented by the Hubbard
repulsion. Most electrons will then be confined within a few cells by jumping 
back and forth between Mn nearest neighbors, unless a coherent motion of 
charges over long distances takes place. However, this latter should determine 
a sudden drop in the dc resistivity below $T_c$, observed in\cite{Tokura}  
Nd$_{0.5}$Sr$_{0.5}$MnO$_3$, not in\cite{Schiffer} 
La$_{0.5}$Ca$_{0.5}$MnO$_3$, where the small size of the Ca ion reduces the 
hybridization between the Mn 3$d$ and O 2$p$ orbitals. 
In La$_{0.5}$Ca$_{0.5}$MnO$_3$, the hopping of polaronic carriers would then
be able to establish a double-exchange mechanism, not effective enough to 
establish ferromagnetism and high dc conductivity in the whole sample.

The above phase-separation scenario for La$_{0.5}$Ca$_{0.5}$MnO$_3$ at  
$T_c > T >T_N$ is consistent with the present spectra at infrared wavelengths, 
where any inhomogeneity on a scale much larger than $L$ is averaged out. 
In Fig. 1(b), the (poorly) mobile holes of the FM regions then appear as they 
were the excited states of the commensurate CDW condensate in the AFM clusters. 
The consistent behaviors of $\chi (T)$, $n_{eff}$ (T), and $\epsilon (T)$ 
between $T_c$ and $T_0$ are easily explained: 
all those quantities in fact measure in different ways the increasing size 
of the AFM clusters at the expense of the FM regions.

\begin{figure}
{\hbox{\psfig{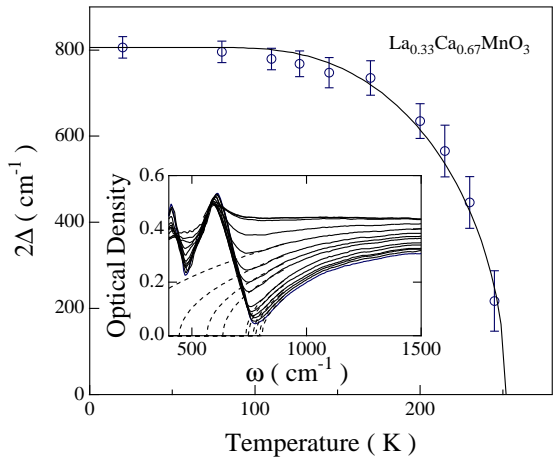}}}
\caption{Width of the optical gap $2\Delta (T)$ in 
La$_{0.33}$Ca$_{0.67}$MnO$_3$, with its fit to Eq. (5) (solid line). The 
spectra used to extract $2\Delta (T)$ by the extrapolations based on Eq. (3) 
(dashed lines) are shown in the inset.}
\label{fig3}
\end{figure} 

However, in order to firmly establish the above interpretation, it remains to 
show that the regions with commensurate CO may produce an infrared gap which 
behaves according to Eq. (5). To this purpose, we show in Fig. 3 the 
midinfrared spectra of La$_{0.33}$Ca$_{0.67}$MnO$_3$. This sample exhibits 
commensurate charge ordering below $T_{CO}$ = 265 K, with wavevector
$\vec q = (2 \pi/a)({1\over 3}, 0, 0)$.\cite{Ramirez} Its absorption spectra, 
reported in the inset of Fig. 3, show formation of a gap in the infrared 
background, starting around $T_{CO}$. No appreciable effects around 
the PM-AFM transition at $T_N \simeq$ 140 K, neither hysteretic phenomena, 
affect $2\Delta (T)$, as obtained by use of Eq. (3). Also in the commensurate 
CDW of $x = 0.67$, then, $2\Delta (T)$  is well fit by Eq. (5). One obtains 
$T_c \simeq$ 255 K (in good agreement with the above $T_{CO}$), $T_0^C = T_0^H$ 
= 35 K, and $2\Delta_0$ = 810 $\pm 25$ cm$^{-1}$. One thus obtains 
$2\Delta_0/k_BT_c \simeq$ 4.6, a value close to that observed in the FM phase 
of La$_{0.5}$Ca$_{0.5}$MnO$_3$. 
 
In conclusion, the present paper first reports on the optical
response of charge density waves interacting with a magnetic background. 
Both the incommensurate
charge ordering in La$_{0.5}$Ca$_{0.5}$MnO$_3$ and the commensurate ordering
in La$_{0.33}$Ca$_{0.67}$MnO$_3$ open in the infrared absorption a gap
well described by a standard BCS-like equation, provided that $T$ = 0 is 
replaced with a finite $T_0<T_N$. The values for $2\Delta_0/k_BT_c$ 
indicate a moderately strong coupling and authorize {\it a posteriori} 
the use of the above equation. These results also suggest an explanation for 
the intriguing coexistence of 
ferromagnetism and charge ordering in La$_{0.5}$Ca$_{0.5}$MnO$_3$. It can be
reconciled with the double-exchange model,
provided that one introduces a phase-separation scenario. Commensurate 
charge-ordered clusters, probably antiferromagnetic, compete 
with disordered ferromagnetic domains where most charges are confined
within a few cells. At the long infrared wavelengths, these 
poorly mobile holes appear as the excited states of the CDW condensate 
confined in the commensurate clusters. 

\acknowledgments
We wish to thank Denis Feinberg and Marco Grilli for many stimulating 
discussions. 


\end{multicols}

\begin{references}
\bibitem{nota} 
The non vanishing dc conductivity of La$_{0.5}$Ca$_{0.5}$MnO$_3$ at 300 K 
($\sigma (0)$ = 100 $\Omega^{-1}$ cm$^{-1}$)\cite{Schiffer}) does not 
contradict the absorption drop observed in Fig. 1(b) for $\omega <$ 
250 cm$^{-1}$. The latter effect is due to the  
penetration depth $\lambda \propto 1/\sqrt{\sigma \omega}$ of the infrared
radiation, which at low $\omega$ becomes comparable with the average dimension 
of the perovskite grains dissolved in CsI.
\end{references}
\end{document}